\documentclass[preprint,showpacs,showkeys,groupedaddress,superscriptaddress]{revtex4}
\def\p{\partial}

\usepackage{epsf,epsfig}
\usepackage{epsf,epsfig}
\usepackage[psamsfonts]{amssymb}
\usepackage{amsmath}
\usepackage{bm}
\begin{document}
\title{Green's Function expansion of scalar and vector fields in the presence of a medium}
\author{Fardin Kheirandish}
\email[]{fardin_kh@phys.ui.ac.ir}
\affiliation{ Department of
Physics, Islamic Azad University, Shahreza Branch, Shahreza,
Iran.}
\author{Shahriar Salimi}
\email[]{shsalimi@uok.ac.ir}
\affiliation{Department of Physics,
Faculty of Science, University of Kurdistan, Sanandaj, Iran.}
\begin{abstract}
\noindent Based on a canonical approach and functional-integration
techniques, a series expansion of Green's function of a scalar
field, in the presence of a medium, is obtained. A series
expansion for Lifshitz-energy, in finite-temperature, in terms of
the susceptibility of the medium is derived and the whole
formalism is generalized to the case of electromagnetic field in
the presence of some dielectrics. A covariant formulation of the
problem is presented.
\end{abstract}
\pacs{05.40.-a, 03.70.+k, 77.22.-d, 81.07.-b} \keywords{Scalar
field; Coupling function; Susceptibility function; Generating
functional} \maketitle
\section{INTRODUCTION}\label{int}
\noindent Quantum field theory is the quantum mechanics of
continuous systems and fully developed in quantum electrodynamics
which is the most successful theory in physics \cite{Greiner}.
Quantum field theory trough path-integrals bridge to statistical
mechanics and its applications include many branches of physics
like, particle physics, condensed-matter physics, atomic physics,
astrophysics and even economics \cite{Kleinert}. Usually we are
interested in a quantum field which has to be considered in the
presence of a matter field described by some bosonic fields. For
example, in quantum optics there are situations where the
electromagnetic field quantization should be achieved in the
presence of a general magnetodielectric medium
\cite{Vogel,kheirandish,Matloob} or in calculating the effect of
matter fields on Casimir forces \cite{Green,K-M-S}. In these cases
the matter field should be included directly into the process of
quantization. Unfortunately there are very few problems where the
interested physical quantities, like for example, the Casimir
force, can be determined analytically and so finding an effective
approximation method is necessary. A fundamental quantity in a
quantum field theory is the propagator or the Green's function
\cite{Greiner-field} from which many physical quantities may be
extracted. Here, using path-integrals, and based on a microscopic
approach, we begin from a Lagrangian and obtain an expansion for
the two-point correlation function i.e., the Green's function in
terms of the susceptibility function of the medium for both scalar
and electromagnetic fields in the presence of an arbitrary linear
magnetodielectric medium. As an example of applications of these
expansions for the case of a real scalar field we have introduced
an expansion for the free energy or Lifshitz energy in the
presence of some arbitrary dielectrics \cite{Lifshitz,Ramin}. Also
we have considered the covariant formulation of the
electromagnetic field in the presence of a linear
magnetodielectric \cite{Amooshahi} which may have applications in
quantum optics or dynamical Casimir effects \cite{Ramin-Faez}.
\section{SCALAR FIELD}
\noindent Let us start the section with a simple but efficient
field theory which have a wide range of applications in many
branches of physics, i.e., the Lagrangian of a real Klein-Gordon
field in $3+1$-dimensional space-time $(x=({\bf x},x^0)\in
\mathbb{R}^{3+1})$, with the following Lagrangian density
\begin{equation}\label{1}
\mathcal{L}_s =
\frac{1}{2}\p_{\mu}\varphi(x)\p^{\mu}\varphi(x)-\frac{1}{2}m^2\varphi^2
(x),
\end{equation}
and let the medium be modeled by a continuum of harmonic
oscillators which is usually called the Hopfield model of a
reservoir \cite{Hopfield}
\begin{equation}\label{2}
\mathcal{L}_m =\frac{1}{2}\int_{0}^{\infty}
d\omega\,\left(\dot{Y}_{\omega}^2(x)-\omega^2
Y_{\omega}^2(x)\right),
\end{equation}
the interaction between the scalar field and its medium is assumed
to be linear and described by
\begin{equation}\label{3}
\mathcal{L}_{int} =\int_{0}^{\infty} d\omega\,f(\omega,{\bf
x})\dot{Y}_{\omega}(x)\varphi(x).
\end{equation}
Having the total Lagrangian we can quantize the system using
path-integral techniques. An important quantity in any field
theory is the generating functional from which $n$-point
correlation functions can be obtained from successive functional
derivatives. Here our purpose is to find two-point correlation
functions or Green's functions in terms of the susceptibility of
the medium. For this purpose let us first find the free generating
functional which can be written as
\begin{eqnarray}\label{4}
W_0[J,\{J_\omega\}]&=&\int D\varphi\, e^{\frac{\imath}{\hbar}\int
d^4\,x\,\{-\frac{1}{2}\varphi[\Box+m^2]\varphi+J\varphi\}}\int
\prod_{\omega}DY_{\omega}\, e^{\frac{\imath}{\hbar}\int d^4
x\,\int_{0}^{\infty}d\omega\{-\frac{1}{2}Y_{\omega}[\p_{t}^2+\omega^2]Y_{\omega}+J_{\omega}Y_{\omega}\}}\nonumber\\
&=& \int D\varphi e^{-\frac{1}{2}\langle\varphi|\hat{A}|\varphi\rangle+\langle
J|\varphi\rangle}\int\prod_{\omega} DY_{\omega} e^{-\frac{1}{2}\int_{0}^{\infty}\{\langle
Y_{\omega}|\hat{B}_\omega|Y_{\omega}\rangle+\langle J_{\omega}|Y_{\omega}\rangle\}}\nonumber\\
\end{eqnarray}
where we have defined
\begin{eqnarray}\label{5}
&&\hat{A}=\frac{\imath}{\hbar}(\Box+m^2),\hspace{1.5cm}\hat{B}_{\omega} = \frac{\imath}{\hbar}(\p_{t}^2+\omega^2),\nonumber\\
&&\rho(x) =
\frac{\imath}{\hbar}J(x),\hspace{1.8cm}\rho_{\omega}(x)=\frac{\imath}{\hbar}J_{\omega}(x).
\end{eqnarray}
Now using the following formula
\begin{equation}\label{6}
\int D\varphi(x)e^{-\frac{1}{2}\langle \varphi
|\hat{A}|\varphi\rangle+\langle\rho|\varphi\rangle}=(det\hat{A})^{-\frac{1}{2}}e^{\frac{1}{2}\langle\rho|\hat{A}^{-1}|\rho\rangle}
\end{equation}
Eq. (\ref{4}) can be rewritten as
\begin{equation}\label{7}
W_0[J,\{J_{\omega}\}]=N e^{\frac{1}{2}\langle\rho|\hat{A}^{-1}|\rho\rangle} e^{\frac{1}{2}\int_{0}^{\infty}
d\omega\,\langle\rho_{\omega}|\hat{B}_{\omega}^{-1}|\rho_{\omega}\rangle}
\end{equation}
where
$N=(det\hat{A})^{-\frac{1}{2}}\prod_{\omega}(det\hat{B}_{\omega})^{-\frac{1}{2}}$
is a renormalization factor. Also from the following definitions
\begin{eqnarray}\label{8}
\frac{\imath}{\hbar}(\Box+m^2)\,G^0 (x,x') &=& \delta^4 (x-x')\nonumber\\
\frac{\imath}{\hbar}(\p_{t}^2+\omega^2)\,G^0_{\omega}(x,x') &=&
\delta^4 (x-x')
\end{eqnarray}
we will find
\begin{eqnarray}\label{9}
\hat{A}^{-1} &=& G^0 (x,x')=\imath\hbar\int\frac{d^4
k}{(2\pi)^4}\frac{e^{\imath
k\cdot(x-x')}}{k^2-m^2}\nonumber\\
\hat{B}_{\omega}^{-1} &=&
G^{0}_{\omega}(x,x')=\imath\hbar\delta^{3}({\bf x}-{\bf
x}')\int\frac{d\,k^0}{2\pi}\frac{e^{\imath k^0
(x^0-x'^{0})}}{(k^{0})^2-\omega^2}
\end{eqnarray}
with the following Fourier transforms
\begin{eqnarray}\label{10}
\tilde{G}^0 (k) &=& \frac{\imath\hbar}{k^2-m^2},\nonumber\\
\tilde{G}^{0}_{\omega}(k^0) &=&
\frac{\imath\hbar}{(k^0)^2-\omega^2}.
\end{eqnarray}
respectively. The free generating functional can now be written as
\begin{equation}\label{11}
W_0[J,\{J_{\omega}\}]=Ne^{-\frac{1}{2\hbar^2}\int d^4 x\int d^4 x'
J(x)G^0(x-x')J(x')} e^{-\frac{1}{2\hbar^2}\int d^4 x\int d^4
x'\int_{0}^{\infty}
J_{\omega}(x)G^{0}_{\omega}(x-x')J_{\omega}(x')},
\end{equation}
and the interacting generating functional can be obtained from the
free generating functional using the following formula
\cite{Greiner-field}
\begin{eqnarray}\label{12}
W[J,\{J_{\omega}\}] &=& e^{\frac{\imath}{\hbar}\int d^4
x\int_{0}^{\infty}
d\omega\,f(\omega,x)\left(\frac{\hbar}{\imath}\frac{\delta}{\delta
J(x)}\right)\frac{\p}{\p
x^{0}}\left(\frac{\hbar}{\imath}\frac{\delta}{\delta J_{\omega}(x)}\right)}W_0[J,\{J_{\omega}\}]\nonumber\\
&=& N\,e^{-\imath\hbar\int_{0}^{\infty} d\omega\int d^4
x\,f(\omega,x)\frac{\delta}{\delta J(x)}\frac{\p}{\p
x^{0}}\frac{\delta}{\delta
J_{\omega}(x)}}\nonumber\\
&\times &\,e^{-\frac{1}{2\hbar^2}\int d^4 x \int d^4 x'
J(x)G^{0}(x-x')J(x')} e^{-\frac{1}{2\hbar^2}\int d^4 x \int d^4 x'
\int_{0}^{\infty}
d\omega\,J_{\omega}(x)G^{0}_{\omega}(x-x')J_{\omega}(x')}\nonumber\\
\end{eqnarray}
Having the generating functional, the two-point function, i.e. the
Green's function can be obtained as
\begin{equation}
G(x,x')=(\frac{\hbar}{\imath})^2\frac{\delta^2}{\delta J(x)\delta J(x')}W[J,\{J_{\omega}\}]\bigg
|_{j,\{j_{\omega}\}=0}.
\end{equation}
Now let us assume that the coupling function between the
Klein-Gordon field and its medium is weak-one can also assume that
the susceptibility of the medium is not far from vacuum- such that
it can be considered as an expansion parameter which can be used
to find a series solution for the Green's function or the
correlation function. Using Eq.(\ref{12}) and after some
straightforward calculations, we find the following expansion for
the Green's function in frequency domain
\begin{eqnarray}\label{23}
&& G({\bf x}-{\bf x'},\omega)= G^{\,0}({\bf x}-{\bf
x'},\omega)+\int_{\Omega}d^{\,3}{\bf z_1}G^{\,0}({\bf x}-{\bf
z_1},\omega)[\omega^2\tilde{\chi}(\omega,{\bf z_1})]G^{\,0}({\bf z_1}-{\bf x'},\omega)+\nonumber\\
&&\int_{\Omega}\int_{\Omega}d^{\,3}{\bf z_1}d^{\,3}{\bf
z_2}\,G^{\,0}({\bf x}-{\bf
z_1},\omega)[\omega^2\tilde{\chi}(\omega,{\bf z_1})]G^{\,0}({\bf
z_1}-{\bf z_2},\omega)[\omega^2\tilde{\chi}(\omega,{\bf
z_2})]G^{\,0}({\bf z_2}-{\bf x'},\omega)+\cdots\nonumber\\
\end{eqnarray}
and it can be easily shown that it satisfies the Green's function
equation which we will find in the next section. Note that since
$\tilde{\chi}(\omega,{\bf x})=0$, for ${\bf x}\notin\Omega$, so we
can rewrite the expansion (\ref{23}) in a more compact or matrix
form as fallows
\begin{eqnarray}
G(\omega)&=&
G^{\,0}(\omega)+G^{\,0}(\omega)\,[\omega^2\,\tilde{\chi}(\omega)]\,G^{\,0}(\omega)+
G^{\,0}(\omega)\,[\omega^2\,\tilde{\chi}(\omega)]G^{\,0}(\omega)[\omega^2\,\tilde{\chi}(\omega)]\,G^{\,0}(\omega)+\cdots\nonumber\\
&=&
G^{\,0}(\omega)\,[\mathbb{I}-\omega^2\tilde{\chi}(\omega)\,G^{\,0}(\omega)]^{-1}.
\end{eqnarray}

\subsection{Equations of motion}
\noindent In this section we find the equations of motion for the
fields and in particular we obtain a Langevin type equation for
the scalar field
\begin{equation}\label{13}
\p_{\mu}\frac{\p\mathcal{L}}{\p(\p_{\mu}\varphi)}=\frac{\p
\mathcal{L}}{\p\varphi}\Longrightarrow(\Box+m^2)\,\varphi=\int_{0}^{\infty}d\omega\,f(\omega,{\bf
x})\dot{Y}_{\omega}(x),
\end{equation}
\begin{equation}\label{14}
\p_{\mu}(\frac{\delta\mathcal{L}}{\delta_{\mu}
Y_{\omega}})=\frac{\delta\mathcal{L}}{\delta
Y_{\omega}}\Longrightarrow\ddot{Y_{\omega}}+\omega^2
Y_{\omega}=-f(\omega,{\bf x})\,\dot{\varphi}(x).
\end{equation}
By solving Eq.(\ref{14}) and inserting it into Eq.(13), we find a
Langevin equation for the Klein-Gordon field
\begin{equation}\label{15}
(\Box+m^2)\,\varphi(x)+\frac{\p}{\p
t}\int_{-\infty}^{t}dt'\,\chi(t-t',{\bf x})\frac{\p}{\p
t'}\,\varphi({\bf x},t')=\xi(x),
\end{equation}
where $\chi(\tau,{\bf x})$ is the susceptibility function or the memory of the medium with the following Fourier
transform
\begin{equation}\label{16}
\tilde{\chi}(\omega,{\bf x})=\int_{0}^{\infty}
d\omega'\,\frac{f^2(\omega',{\bf x})}{\omega'^2-\omega^2+\imath
0^+}.
\end{equation}
The source field $\xi(x)$ is defined by
\begin{equation}\label{17}
\xi(x)=\int_{0}^{\infty}d\omega\,f(\omega,{\bf x})\dot{Y}_{\omega}^{N}(x)
\end{equation}
where
\begin{equation}\label{18}
Y_{\omega}^{N}({\bf x},t)=\cos(\omega t)Y_{\omega}({\bf
x},0)+\frac{\sin(\omega t)}{\omega}\dot{Y}_{\omega}({\bf x},0).
\end{equation}
From Eqs.(\ref{17},\ref{18}), we see that the source field depends
on initial values of the reservoir fields so it can be considered
as a noise field. The Green's function of Eq.(\ref{15}) satisfies
\begin{equation}\label{19}
(\Box+m^2)\,G({\bf x}-{\bf x'},t-t')+\frac{\p}{\p
t}\int_{-\infty}^{t}dt''\,\chi(t-t'',{\bf x})\frac{\p}{\p
t''}G({\bf x}-{\bf x'},t''-t')=\delta({\bf x}-{\bf x'},t-t').
\end{equation}
In a homogeneous medium, where the memory function is position
independent, Eq.(\ref{19}) can be solved easily in reciprocal
space
\begin{eqnarray}\label{20}
\tilde{G}({\bf k},\omega) &=& \frac{1}{{\bf k}^2-\omega^2+m^2-\omega^2\tilde{\chi}(\omega)}=\frac{1}{{\bf
k}^2-\omega^2+m^2-\int d\omega'\,\frac{\omega^2\,f^2(\omega')}{\omega'^2-\omega^2+\imath 0^+}}\nonumber\\
&=& \frac{1}{{\bf k}^2-\omega^2\epsilon(\omega)+m^2}
\end{eqnarray}
where $\epsilon(\omega)=1+\tilde{\chi}(\omega)$ can be considered
as the dielectric function corresponding to the medium. From
Eq.(\ref{20}) it is clear that the Green's function in the
presence of a homogeneous medium can be obtained from the Green's
function of the free space simply by substituting $\omega^2$ with
$\omega^2\epsilon(\omega)$. Eq.(\ref{15}) in frequency-space can
be written as
\begin{equation}\label{21}
(-\nabla^2-\omega^2\epsilon(\omega,{\bf x})+m^2)\,G({\bf x}-{\bf x'},\omega)=\delta({\bf x}-{\bf x'})
\end{equation}
In some simple geometries the dielectric function $\epsilon(\omega,{\bf x})$ depends on ${\bf x}$ as follows
\begin{equation}\label{22}
\epsilon(\omega,{\bf x})=\left\{\begin{array}{ll}
\epsilon(\omega) & \textrm{if ${\bf x}\in\Omega$}\\
1 & \textrm{if ${\bf x}\notin\Omega$}
\end{array}\right.
\end{equation}
where $\Omega$ is a region or the union of regions where the space
is filled with a homogeneous but frequency dependent medium with
the dielectric function $\epsilon(\omega)$. In this case the
Green's function can be found in some regular geometries
\cite{Green} but for an arbitrary dielectric function it is quite
complicated and in this case we find a series solution in terms of
free Green's function and susceptibility of the medium. Note that
in some geometries, electromagnetic field can be considered as two
massless Klein-Gordon fields, and the scalar formalism can help
for example in obtaining Lifshitz energies or Casimir forces in
such geometries \cite{K-M-S}.

\subsection{Partition function}
\noindent Having the expansion (\ref{23}) let us find the
partition function in the presence of some dielectrics defined by
the dielectric function $\epsilon(\omega, {\bf x})$ which as a
special case may be given by (\ref{22}). The partition function of
a real scalar field in the presence of a medium according to the
modified Green's function given by (\ref{21}) can be written as
\begin{equation}\label{24}
\Xi=\int\,D\varphi\,e^{\frac{\imath}{\hbar}S}=\int\,D\varphi\,e^{\frac{\imath}{\hbar}\int\,d^4x\,\mathcal{L}}
\end{equation}
where $\mathcal{L}$ is given by (\ref{1}). The partition function
in frequency domain can be written as \cite{Kapusta}
\begin{equation}\label{25}
\Xi=\int\,D\varphi\,e^{-\frac{\imath}{2\hbar}\int\frac{d\,\omega}{2\pi}\,\int
d^3{\bf x}\,\tilde{\varphi}({\bf
x},-\omega)[-\omega^2\epsilon(\omega,{\bf
x})-\nabla^2+m^2]\tilde{\varphi}({\bf x},\omega)}.
\end{equation}
If we make a Wick rotation $\omega=\imath\nu$ in frequency domain
the action will be Euclidean and the free energy can be determined
from $E=-\frac{\hbar}{\tau}\ln \Xi$, where $\tau$ is the duration
of interaction which is taken to be sufficiently large. Using
standard path-integral techniques we will find the free energy in
finite temperature $T$ as
\begin{equation}\label{2.26}
E=k_B T\sum_{l=0}^{\infty \,\,\prime}\ln\, \det [\hat{K}(\nu_l;
{\bf x},{\bf x'}]
\end{equation}
where $\nu_l=2\pi l k_b T/\hbar$ is the Matsubara frequency, $k_B$
is the Boltzman constant and the prime over the summation means
that the term corresponding to $l=0$, should be given a half
weight. The kernel $\hat{K}(\nu_l; {\bf r},{\bf
r'})=[\nu_{l}^2\epsilon(\imath\nu_l,{\bf
r})-\nabla^2]\delta^3({\bf r}-{\bf r'})$. Using the identity,
$\ln\, \det [\hat{K}]=tr\,\ln [\hat{K}]$ and the fact that
$\hat{K}(\nu_l; {\bf r},{\bf r'})=G^{-1}(\imath\nu_l;{\bf r},{\bf
r'})$ we find
\begin{equation}\label{2.27}
E=-k_B T\sum_{l=0}^{\infty \,\,\prime}\,tr\,\ln [G(\imath\nu_l;
{\bf r},{\bf r'}]
\end{equation}
now using the expansion ({\ref{23}), we find the following
expansion for free energy in terms of the susceptibility
\begin{equation}\label{2.28}
E=k_B T\sum_{l=0}^{\infty
\,\,\prime}\sum_{n=1}^{\infty}\frac{(-1)^{n+1}}{n}\int\,d^3{\bf
r}_1\cdots d^3{\bf r}_n\,G^{0}(\imath\nu_l;{\bf r}_1-{\bf
r}_2)\cdots\,G^{0}(\imath\nu_l;{\bf r}_n-{\bf
r}_1)\,\chi(\imath\nu_l,{\bf r}_1)\cdots\chi(\imath\nu_l,{\bf
r}_n)
\end{equation}
where $G^{0}({\bf r}-{\bf r'};\imath\nu_l)$ is given by
\begin{equation}\label{2.29}
G^{0}(\imath\nu_l;{\bf r}-{\bf
r'})=\frac{1}{4\pi}\frac{e^{-\sqrt{m^2+\nu_l^2}\,|{\bf r}-{\bf
r'}|}}{|{\bf r}-{\bf r'}|}
\end{equation}
which corresponds to a Yukawa potential with the modified mass
$\sqrt{m^2+\nu_l^2}$.
\section{ELECTROMAGNETIC FIELD}
\noindent In this section we use the Coulomb gauge i.e.
$\nabla\cdot {\bf A}=0$, $A^{0}=0$ and find a similar expansion
for the Green's function of the electromagnetic field in the
presence of some arbitrary regions of matter which as an example
can have applications in calculating the Casimir forces. For this
purpose let us take the total Lagrangian density as follows
\cite{Kheirandish-Soltani}
\begin{equation}\label{24}
\mathcal{L}=\frac{1}{2}\epsilon_0({\bf E}^2-\frac{1}{\mu_0}{\bf
B}^2)+\frac{1}{2}\int_{0}^{\infty}d\omega(\dot{{\bf Y_{\omega}}}^2(x)-\omega^2{\bf Y_{\omega}}^2(x))+\int
d\omega f(\omega,{\bf x}){\bf A}\cdot\dot{{\bf Y_{\omega}}}
\end{equation}
The interacting generating functional can be written in terms of the vector potential and the medium fields as
\begin{eqnarray}
W &=& \int D[{\bf A}]\prod_{\omega}D[{\bf
Y}_{\omega}]\exp\frac{\imath}{\hbar}\int d^4 x
\bigg\{-\frac{1}{2}A_i
\hat{K}_{ij}A_j-\int_{0}^{\infty}\frac{1}{2}Y_{\omega,\,i}(\partial_t^2
+\omega^2)\,\delta_{ij}\,Y_{\omega,j}\nonumber\\
&+& \int_{0}^{\infty} d\omega f(\omega,{\bf x})A_i
\dot{Y}_{\omega,i}+ J_i A_i +\int_{0}^{\infty} d\omega\,
J_{\omega,\,i}Y_{\omega,\,i}\bigg\}
\end{eqnarray}
where summation over repeated indices is assumed and the kernel $\hat{K}_{ij}$ is defined by
\begin{equation}
\hat{K}_{ij}=\left[\epsilon_0\partial_0^2-\frac{1}{\mu_0}\nabla^2\right]\delta_{ij}+\frac{1}{\mu_0}\partial_i\partial_j
\end{equation}
Now from the equation
\begin{equation}
G_{ij}(x,x')=(\frac{\hbar}{\imath})^2\frac{\delta^2}{\delta J_i(x)\delta J_j(x')}W[J,\{J_{\omega}\}]\bigg
|_{j,\{j_{\omega}\}=0}.
\end{equation}
and similar calculations we will find the following expansion for the Green's function in frequency domain
\begin{eqnarray}\label{25}
&& G_{ij}({\bf x}-{\bf x'},\omega)= G^{\,0}_{ij}({\bf x}-{\bf
x'},\omega)+\int_{\Omega}d^3\,{\bf z_1}G^{\,0}_{il}({\bf x}-{\bf
z_1},\omega)[\omega^2\tilde{\chi}(\omega,{\bf z_1})]G^{\,0}_{lj}({\bf z_1}-{\bf x'},\omega)+\nonumber\\
&&\int_{\Omega}\int_{\Omega}d^3{\bf z_1}d^3{\bf
z_2}\,G^{\,0}_{il}({\bf x}-{\bf
z_1},\omega)[\omega^2\tilde{\chi}(\omega,{\bf z_1})]G^{\,0}_{l
m}({\bf z_1}-{\bf
z_2},\omega)[\omega^2\tilde{\chi}(\omega,{\bf z_2})]G^{\,0}_{mj}({\bf z_2}-{\bf x'},\omega)+\cdots\nonumber\\
\end{eqnarray}
which in matrix form can be written as
\begin{eqnarray}
\mathbb{G}(\omega)&=&
\mathbb{G}^{\,0}(\omega)+\mathbb{G}^{\,0}(\omega)[\omega^2\,\tilde{\chi}(\omega)]\mathbb{G}^{\,0}(\omega)+
\mathbb{G}^{\,0}(\omega)[\omega^2\,\tilde{\chi}(\omega)]\mathbb{G}^{\,0}(\omega)
[\omega^2\,\tilde{\chi}(\omega)]\mathbb{G}^{\,0}(\omega)+\cdots\nonumber\\
&=&
\mathbb{G}^{\,0}(\omega)\left[\mathbb{I}-\omega^2\,\tilde{\chi}(\omega)\,\mathbb{G}^{\,0}(\omega)\right]^{-1}.
\end{eqnarray}
\subsection{Equations of motion}
\noindent From Lagrangian density (\ref{2}) we find the following
equations
\begin{equation}\label{26}
\hat{K}_{ij}A_j = \int_{0}^{\infty}d\omega\,f(\omega,{\bf x})\dot{Y}_{\omega,i}
\end{equation}
\begin{equation}\label{27}
\ddot{Y}_{\omega,i}+\omega^2 Y_{\omega,i}=-f(\omega,{\bf x})\dot{A}_i
\end{equation}
Solving Eq.(\ref{27}) and inserting the solution into
Eq.(\ref{26}) we find
\begin{equation}\label{28}
\hat{K}_{ij}A_j+\frac{\partial}{\partial t}\int_{0}^{\infty} d\omega\,f^2(\omega,{\bf x})\int
dt'\,G_{\omega}(t-t')\frac{\partial}{\partial t'}A_i (t')=\int_{0}^{\infty} d\omega\,f(\omega,{\bf
x})\dot{Y}^{N}_{\omega,i}
\end{equation}
where
\begin{equation}\label{29}
G_{\omega}(t-t')=\int \frac{d\omega'}{2\pi}\,\frac{e^{\imath\omega'(t-t')}}{\omega^2-\omega'^2}
\end{equation}
and $Y^{N}_{\omega,i}$ is the homogeneous solution of
Eq.(\ref{27}) which depends on the initial values of the medium
fields and can be considered as a noise or fluctuating field which
does not affect the Green's function. Using Eqs.(\ref{17}) and
(\ref{29}) we can rewrite Eq.(28) as
\begin{equation}\label{30}
\hat{K}_{ij}A_j+\frac{\partial}{\partial t}\int_{0}^{\infty} \frac{d\omega'}{2\pi}\,\tilde{\chi}(\omega',{\bf
x})\int dt'\,e^{\imath\omega'(t-t')}\frac{\partial}{\partial t'}A_i (t')=\int_{0}^{\infty}
d\omega\,f(\omega,{\bf x})\dot{Y}^{N}_{\omega,i}
\end{equation}
which in frequency-domain can be written as
\begin{equation}\label{31}
\left[(-\epsilon_0
\omega^2-\frac{1}{\mu_0}\nabla^2)\,\delta_{ij}+\frac{1}{\mu_0}\partial_i\partial_j\right]\tilde{A}_j({\bf
x},\omega)-\omega^2\tilde{\chi}(\omega,{\bf x})\tilde{A}_i
(\omega,{\bf x})=\int_{0}^{\infty}
d\omega\,\imath\omega\,f(\omega,{\bf
x})\tilde{Y}^{N}_{\omega,i}(\omega,{\bf x})
\end{equation}
The Green's function of Eq.(\ref{31}) satisfies
\begin{equation}\label{32}
\left[(-\epsilon_0 (1+\tilde{\chi}(\omega,{\bf
x})\,\omega^2\,\delta_{ij})-\frac{1}{\mu_0}\nabla^2)\,\delta_{ij}+\frac{1}{\mu_0}\partial_i\partial_j\right]\,G_{jk}({\bf
x},{\bf x'},\omega)=\frac{1}{\mu_0}\delta^3({\bf x}-{\bf
x'})\delta_{ij}
\end{equation}
which using the definitions $\epsilon(\omega,{\bf x})=\epsilon_0
[1+\tilde{\chi}(\omega,{\bf x}]$ and $c^{-2}=\epsilon_0 \mu_0$ can
be written as
\begin{equation}\label{33}
\left[-\frac{\omega^2}{c^2}\epsilon(\omega,{\bf
x})\delta_{ij}-\nabla^2\delta_{ij}+\partial_i\partial_j\right]\,G_{jk}({\bf
x},{\bf x'},\omega)=\delta^3({\bf x}-{\bf x'})\delta_{ik},
\end{equation}
and it can be easily shown that the Green's function (\ref{25})
satisfies Eq.(\ref{33}).

A similar approach can be followed to find the partition function
in terms of the susceptibility of the medium as follows
\begin{eqnarray}\label{2.28}
E &=& k_B T\sum_{l=0}^{\infty
\,\,\prime}\sum_{n=1}^{\infty}\frac{(-1)^{n+1}}{n}\int\,d^3{\bf
x}_1\cdots\int\,d^3{\bf x}_n\,G^{0}_{i_1 i_2}({\bf x}_1-{\bf
x}_2;\imath\nu_l)\cdots\,G^{0}_{i_n i_1}({\bf x}_n-{\bf
x}_1;\imath\nu_l)\nonumber\\
& \times & \,\chi(\imath\nu_l,{\bf
x}_1)\cdots\chi(\imath\nu_l,{\bf x}_n),
\end{eqnarray}
where the free Green's function $G^{0}_{i j}({\bf x}-{\bf
x'};\imath\nu_l)$ satisfies Eq.(\ref{33}) with
$\epsilon(\omega,{\bf x})=1$ and $\omega=\imath\nu_l$. By defining
${\bf r}={\bf x}-{\bf x'}$, we find
\begin{equation}
G^{0}_{i j}({\bf
r};\imath\nu_l)=\frac{\nu_l^2}{c^2}\frac{e^{-\frac{\nu_l
r}{c}}}{4\pi r}\left[\delta_{i j}(1+\frac{c}{\nu_l
r}+\frac{c^2}{\nu_l^2 r|^2})-\frac{r_i r_j}{r^2}(1+\frac{3c}{\nu_l
r}+\frac{3c^2}{\nu_l^2 r^2})\right]+\frac{1}{3}\delta_{ij}\delta^3
({\bf r}).
\end{equation}
In zero temperature, the summation over the positive integer $l$
is replaced by an integral according to the rule $\hbar
\displaystyle\int_0^\infty {\frac{{d\nu }}{{2\pi }}} f(\iota \nu
)\, \to \,k_B T\sum\limits_{l = 0}^{\infty\,\prime} {f(\iota \nu_l
)}$. For a nice discussion of Casimir-Lifshitz interaction between
dielectrics of arbitrary geometry see \cite{Ramin}.
\subsection{Covariant formulation}
\noindent In reference \cite{Amooshahi} electromagnetic field
quantization in a moving medium has been investigated by
considering the medium to be modeled by a continuum of tensor
fields. But the medium can also be modeled by a continuum of
scalar fields, i.e., Klein-Gordon fields which we will follow
here. So we consider the electromagnetic field interacting with a
moving medium, a situation which can have applications in dynamic
Casimir effects. For this purpose we consider the following
Lorentz invariant Lagrangian density
\begin{equation}\label{34}
\mathcal{L}(x)=\frac{1}{2}\p_{\mu}A_{\nu}\p^{\,\mu}A^{\nu}+\frac{1}{2}\int_{0}^{\infty}
d\omega\,[\p_{\mu}Y_{\omega}\p^{\,\mu}Y_{\omega}-\omega^2
Y_{\omega}^2]+\int_{0}^{\infty} d\omega\,f^{\mu\nu}(\omega,{\bf
x})\,Y_{\omega}\p_{\mu}A_{\nu}.
\end{equation}
where $f^{\mu\nu}(\omega,{\bf x})$ is an antisymmetric coupling
tensor which couples the electromagnetic field to its medium and
is related to the susceptibility of the medium trough
Eq.(\ref{suscep}). From Euler-Lagrange equations we find
\begin{equation}\label{35}
\Box\, A^{\nu} = -\p_{\mu}K^{\mu\nu}(x),
\end{equation}
\begin{equation}\label{36}
(\Box +\omega^2)\,Y_{\omega}(x) = f^{\mu\nu}(\omega,x)\p_{\mu} A_{\nu}(x)
\end{equation}
where $\Box=\p_t^{\,2}-\nabla^2$, and the antisymmetric tensor
$K^{\mu\nu}(x)=\displaystyle\int_{0}^{\infty}
d\omega\,f^{\mu\nu}(\omega,x)Y_{\omega}$ can be considered as the
polarization tensor of the medium. By solving Eq.(\ref{36}) and
inserting it into Eq.(\ref{35}) we find
\begin{equation}
\Box\,A^{\mu}(x)-\int
d^4x'\,\p_{\nu}\,\p^{\,\prime}_{\alpha}\,\chi^{\,\nu\mu\alpha\beta}(x,x')A_{\beta}(x')=-\p_{\eta}K^{N,\eta\mu}
\end{equation}
where $K^{N,\mu\nu}=\int_{0}^{\infty}
d\omega\,f^{\mu\nu}(\omega,x)Y^{N}_{\omega}$, and $Y^{N}_{\omega}$
is the homogeneous solution of Eq.(\ref{36}) which can be
considered as a noise field. The susceptibility tensor
$\chi^{\nu\mu\alpha\beta}(x,x')$ is defined by
\begin{equation}\label{suscep}
\chi^{\,\nu\mu\alpha\beta}(x,x')=\int_{0}^{\infty}d\omega\,f^{\mu\nu}(\omega,x)\,G^{0}_{\omega}(x-x')
f^{\alpha\beta}(\omega,x')
\end{equation}
where $G^{0}_{\omega}(x-x')$ is given by Eq.(\ref{9}). The Green's
function in this case satisfies the following equation
\begin{equation}
\Box\,G_{\mu\nu}(x-x')-\int d^{\,4}
x''\,g_{\mu\delta}\,\p_{\gamma}\,\p_{\alpha}^{''}\,\chi^{\gamma\delta\alpha\beta}(x,x'')\,G_{\beta\nu}(x''-x)
=g_{\mu\nu}\,\delta^4(x-x')
\end{equation}
and for the Green's function we find the following expansion in
terms of the susceptibility tensor
\begin{eqnarray}
&& G_{\mu\nu}(x,x')=G^{0}_{\mu\nu}(x-x')+\int d^{\,4}z_1\,d^{\,4}
z_2\,G^{0}_{\mu\nu_1}(x-z_1)\,\Gamma^{\nu_1\nu_2}(z_1,z_2)\,G^{0}_{\nu_2\nu}(z_2-x')+\nonumber\\
&& \int d^{\,4} z_1\cdots d^{\,4}
z_4\,G^{0}_{\mu\nu_1}(x-z_1)\,\Gamma^{\nu_1\nu_2}(z_1,z_2)\,
G^{0}_{\nu_2\nu_3}(z_2-z_3)\,\Gamma^{\nu_3\nu_4}(z_3,z_4)\,G^{0}_{\nu_4\nu}(z_4-x')+\cdots\nonumber
\end{eqnarray}
where for simplicity we have defined
$\Gamma^{\nu_1\nu_2}(z_1,z_2)=\p_{\mu_1}\p_{\mu_2}\chi^{\mu_1\nu_1\mu_2\nu_2}(z_1,z_2)$.
So given the susceptibility tensor of the medium, one can find the
Green's function perturbatively in terms of the susceptibility.
Having the Green's function-at least perturbatively- we can
investigate for example the dynamical energy configurations, which
is closely related to the problems of dynamical Casimir effect,
which is under consideration.

\section{Conclusion}
\noindent Based on a Lagrangian approach, scalar and vector field
theory in the presence of a medium, modeled by a continuum of
Klein-Gordon fields, considered and a series expansion for the
Green's function of the theory in terms of the susceptibility of
the medium obtained. From the partition function of the scalar
field, an expression for the free energy in terms of the
susceptibility of the medium obtained and the formalism
generalized to the case of electromagnetic field in the presence
of some dielectrics. Also, the covariant form of the
electromagnetic field in the presence of moving media investigated
and an expression for the Green's function in terms of the
susceptibility tensor obtained which can have applications in
dynamical Casimir effects.

\begin{acknowledgments}
\noindent The authors wish to thank the Office of Research of
Islamic Azad University, Shahreza-Branch, for their support.
\end{acknowledgments}

\end{document}